\documentclass[sigplan,natbib,nonacm]{acmart}
\usepackage{natbib}
\usepackage{amsfonts}
\usepackage{amsmath}
\usepackage{tabulary}
\usepackage{booktabs}
\usepackage{tikz}
\usetikzlibrary{arrows.meta}
\usetikzlibrary{bending}
\usetikzlibrary{positioning}
\usetikzlibrary{calc}
\usetikzlibrary{decorations.text}
\usetikzlibrary{backgrounds}
\usetikzlibrary{automata}

\title{Improvements to \texttt{ltlsynt}}

\author{Florian Renkin}
\orcid{0000-0002-5066-1726}
\affiliation{\institution{LRDE/EPITA}\city{Le Kremlin-Bicêtre}\country{France}}
\email{renkin@lrde.epita.fr}
\author{Philipp Schlehuber}
\affiliation{\institution{LRDE/EPITA}\city{Le Kremlin-Bicêtre}\country{France}}
\email{philipp@lrde.epita.fr}
\author{Alexandre Duret-Lutz}
\orcid{0000-0002-6623-2512}
\affiliation{\institution{LRDE/EPITA}\city{Le Kremlin-Bicêtre}\country{France}}
\email{adl@lrde.epita.fr}
\author{Adrien Pommellet}
\orcid{0000-0001-5530-152X}
\affiliation{\institution{LRDE/EPITA}\city{Le Kremlin-Bicêtre}\country{France}}
\email{adrien@lrde.epita.fr}

\newcommand{\F}{\mathsf{F}} 
\newcommand{\LTLG}{\mathsf{G}} 

\tikzset{
  automaton/.style={
    semithick,shorten >=1pt,
    node distance=1.5cm,
    initial text=,
    every initial by arrow/.style={every node/.style={inner sep=0pt}},
    every state/.style={
      align=center,
      fill=white,
      minimum size=7.5mm,
      inner sep=0pt,
      execute at begin node=\strut,
    }},
  smallautomaton/.style={automaton,
                         node distance=7mm,
                         every state/.style={minimum size=4mm,
                           fill=white,
                           inner sep=1.5pt}},
  >={Stealth[round,bend]},
}

\begin{document}
\begin{abstract}
  We summarize \texttt{ltlsynt}'s evolution since 2018.
\end{abstract}

\maketitle

\section{Introduction and History}

The tool \texttt{ltlsynt}, distributed in the Spot
library~\cite{duret.16.atva2} since version 2.5, was originally
developed by Thibaud Michaud and Maximilien Colange.  They submitted
it to the 2017 and 2018~\cite{michaud.18.synt} editions of SYNTCOMP.
This short document summarizes the improvements brought to
\texttt{ltlsynt} since then.

While both original authors left the project mid-2018 
Maximilien had started working on an alternative approach called LAR
(described below) that was eventually included in the Spot 2.7
release.  Without any submission of \texttt{ltlsynt} to SYNTCOMP'19,
the organizers installed the latest version distributed with Spot
2.7.4 themselves, and uncovered a bug caused by an incorrect
optimization in the LAR approach.


This optimization was reverted in Spot 2.8, and we started
working on an optimized reimplementation of LAR for Spot
2.9, adding other improvement to \texttt{ltlsynt} along the way.

A quick summary of all versions submitted to SYNTCOMP over the years
is given in Table~\ref{tab:summary}.  Note that since the release
calendar of Spot is not aligned with SYNTCOMP, most submissions are
development version containing unreleased patches applied to a
previous release.

\begin{table*}
\begin{tabulary}{\textwidth}{llJ}
  Year & Spot version  & Main news in \texttt{ltlsynt} \\
  \midrule
  2017 & pre-2.4? + patches & first implementation \\
  2018 & 2.5.3 + patches & optimizations to determ., and game solving; incr. determ. approach \\
  2019 & 2.7.4 & (bogus) LAR; improved LTL translation; incr. determ. removed \\
  2020 & 2.9 + patches & reimplemented LAR, split, and game solving; parity minimization \\
  2021 & 2.9.7 + patches & input decomposition; strategy simplification; specialized strategy construction  \\
\end{tabulary}
  \caption{Versions of Spot on which \texttt{ltlsynt} submissions to
    SYNTCOMP were based.\label{tab:summary}}
\end{table*}



\section{Technical Details}

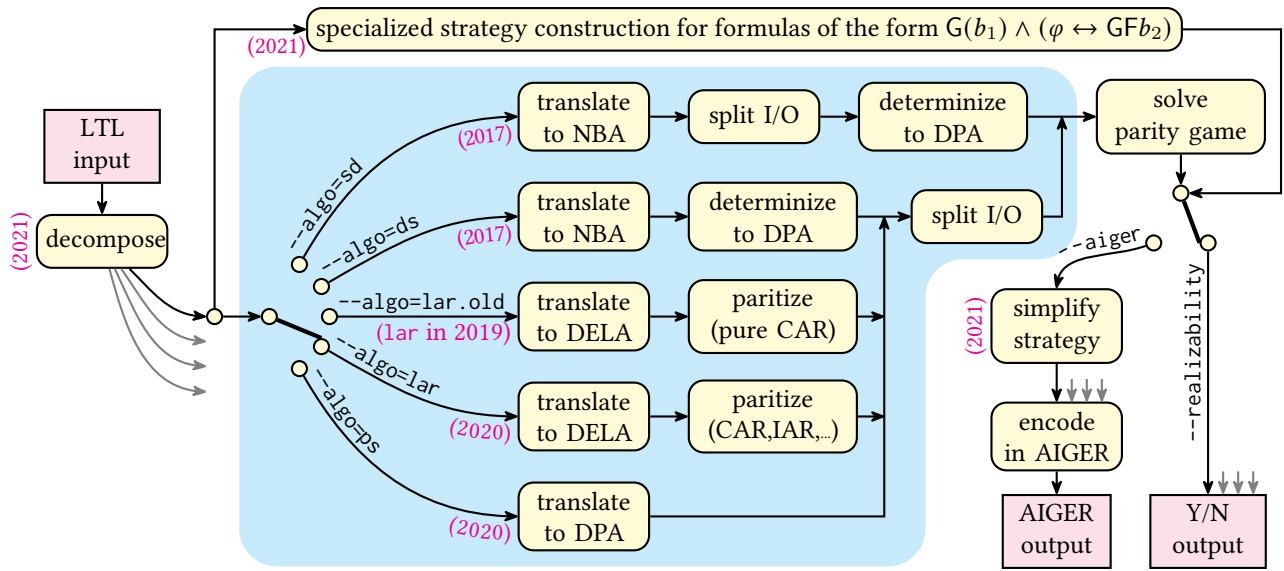
\begin{figure*}[tb]
  \begin{tikzpicture}[thick,
                      node distance=4mm and 5mm,
                      data/.style={text width=1.3cm,minimum height=2em,align=center,draw,fill=magenta!15},
                      wproc/.style={fill=yellow!20,align=center,draw,rounded corners=2mm},
                      proc/.style={wproc,text width=1.5cm,minimum height=2em},
                      lproc/.style={proc,text width=2cm},
                      choice/.style={circle,fill=yellow!20,inner sep=0,minimum size=2mm,draw},
                      algoopt/.style={postaction={decorate,decoration={text along path, raise=1mm,text={|\ttfamily\small|#1}}}},
                      algodwn/.style={postaction={decorate,decoration={text along path, text align=right,raise=-.9em,text={|\small\color{magenta}|#1}}}},
                      outopt/.style={postaction={decorate,decoration={text along path, text align=right,raise=1mm,text={|\ttfamily\small|#1}}}},
                      ]
    \node[proc] (transSD) {translate to NBA};
    \node[proc,right=of transSD] (splitSD) {split I/O};
    \node[lproc,right=of splitSD] (detSD) {determinize to DPA};
    \draw[->] (transSD) edge (splitSD)
              (splitSD) edge (detSD);
    \node[proc,below=of transSD] (transDS) {translate to NBA};
    \node[lproc,right=of transDS] (detDS) {determinize to DPA};
    \node[proc,right=of detDS,xshift=2mm] (splitDS) {split I/O};
    \draw[->] (transDS) edge (detDS)
              (detDS) edge coordinate(middetsplit) (splitDS);
    \node[proc,below=of transDS] (transLARold) {translate to DELA};
    \node[lproc,right=of transLARold] (paritizeLARold) {paritize (pure CAR)};
    \draw[->] (transLARold) edge (paritizeLARold)
              (paritizeLARold) edge (middetsplit |- paritizeLARold);
    \node[proc,below=of transLARold] (transLAR) {translate to DELA};
    \node[lproc,right=of transLAR] (paritizeLAR) {paritize (CAR,IAR,{\tiny ...})};
    \draw[->] (transLAR) edge (paritizeLAR)
              (paritizeLAR) edge (middetsplit |- paritizeLAR);
    \node[proc,below=of transLAR] (transPS) {translate to DPA};
    \draw[->] (transPS) -| (middetsplit);
    \coordinate (choicecenter) at ($(current bounding box.west) - (3.3cm,0)$);
    \path[draw] (choicecenter) node[choice](algoin){}
          +(60:8mm) node[choice](algosd){}
          +(30:8mm) node[choice](algods){}
          +(0:8mm) node[choice](algolarold){}
          +(-30:8mm) node[choice](algolar){}
          +(-60:8mm) node[choice](algops){};

    \draw[->,algoopt={-{}-algo=sd}] (algosd) to[out=60,in=180] (transSD);
    \path[->,algodwn={(2017)~}] (algosd) to[out=60,in=180] (transSD);
    \draw[->,algoopt={-{}-algo=ds}] (algods) to[out=30,in=180] (transDS);
    \path[->,algodwn={(2017)~}] (algods) to[out=30,in=180] (transDS);
    \draw[->,algoopt={-{}-algo=lar.old}] (algolarold) -- (transLARold);
    \path[->,algodwn={(|+\ttfamily|lar|\color{magenta}| in 2019)~}] (algolarold) -- (transLARold);
    \draw[->,algoopt={-{}-algo=lar}] (algolar) to[out=-30,in=180] (transLAR);
    \path[->,algodwn={(2020)~}] (algolar) to[out=-30,in=180] (transLAR);
    \draw[->,algoopt={-{}-algo=ps}] (algops) to[out=-60,in=180] (transPS);
    \path[->,algodwn={(2020)~}] (algops) to[out=-60,in=180] (transPS);
    \draw[ultra thick] (algoin) -- (algolar.north);

    \node[choice,left=of algoin] (bypassin) {};
    \node[proc,left=of bypassin,yshift=1cm] (decomp) {decompose};
    \node[data,above=of decomp] (input) {LTL input};
    \draw[->] (input) -- (decomp);
    \draw[->] (decomp) to[out=-45,in=180] (bypassin);
    \draw[->,gray] (decomp) to[out=-55,in=180] ($(bypassin.west)+(0,-0.33)$);
    \draw[->,gray] (decomp) to[out=-65,in=180] ($(bypassin.west)+(0,-0.66)$);
    \draw[->,gray] (decomp) to[out=-75,in=180] ($(bypassin.west)+(0,-1)$);
    \draw[->] (bypassin) -- (algoin);

    \node[lproc, right=of detSD, xshift=4mm] (solve) {solve\linebreak parity game};

    \node[below=of solve,choice] (reachoice) {};
    \node[below right=of reachoice,choice,xshift=-3mm,yshift=-1mm] (realize) {};
    \node[below left=of reachoice,choice,xshift=3mm,yshift=-1mm] (aiger) {};
    \draw[ultra thick] (reachoice) -- (realize.west);
    \draw[->] (solve) -- (reachoice);

    \node[data] (yesno) at ($(transPS -| realize)+(0,-2mm)$) {Y/N output};
    \node[data, text width=1.2cm, left=of yesno] (output) {AIGER output};
    \node[proc, above=of output,yshift=-1mm] (encode) {encode in AIGER};
    \node[proc, above=of encode,yshift=+1mm] (minimize) {simplify\linebreak strategy};

    \draw[->] (detSD) -- coordinate(middetsolve) (solve);
    \draw[->] (splitDS) -| (middetsolve);

    \draw[<-,outopt={-{}-aiger}] (minimize) to[out=90,in=-160]  (aiger);
    \draw[->] (minimize) -- (encode);
    \draw[gray,->] ($(encode.north)+(2mm,3mm)$) -- ($(encode.north)+(2mm,0)$);
    \draw[gray,->] ($(encode.north)+(4mm,3mm)$) -- ($(encode.north)+(4mm,0)$);
    \draw[gray,->] ($(encode.north)+(6mm,3mm)$) -- ($(encode.north)+(6mm,0)$);
    \draw[<-,outopt={-{}-realizability~}] (yesno) -- (realize);
    \draw[gray,->] ($(yesno.north)+(2mm,3mm)$) -- ($(yesno.north)+(2mm,0)$);
    \draw[gray,->] ($(yesno.north)+(4mm,3mm)$) -- ($(yesno.north)+(4mm,0)$);
    \draw[gray,->] ($(yesno.north)+(6mm,3mm)$) -- ($(yesno.north)+(6mm,0)$);

    \draw[->] (encode) -- (output);

  \begin{scope}[on background layer,overlay]
    \fill[cyan!20,rounded corners=5mm]
    ($(algoin |- transPS.south)+(-.4,-.2)$) |-
    ($(detSD.north -| middetsolve)+(.2,.2)$) |-
    ($(splitDS.south west)+(.2,-.2)$) |- cycle;
  \end{scope}

  \coordinate (solveright) at ($(solve.east)+(2mm,0)$);
  \path (bypassin) -- coordinate(medblue) (solveright);
  \node[wproc,above,xshift=-1mm,yshift=4mm] (bypass) at (medblue |- detSD.north) {specialized strategy construction for formulas of the form $\LTLG(b_1) \land (\varphi \leftrightarrow \LTLG\F b_2)$};
  \draw[->,algodwn={(2021)}] (bypassin) |- (bypass);
  \draw[->] (bypass) -| (solveright) |- (reachoice);
  \node[rotate=90,above,inner sep=1pt] at (minimize.west){\small\textcolor{magenta}{(2021)}};
  \node[rotate=90,above,inner sep=1pt] at (decomp.west){\small\textcolor{magenta}{(2021)}};
  \end{tikzpicture}
  \caption{Architecture of \texttt{ltlsynt}.  The blue zone shows different pipelines for building a parity game, selected by option \texttt{-{}-algo}.  For some types a formulas, a strategy can be constructed directly from a DBA, bypassing the game construction.  If the input is decomposed in multiple conjuncts, recomposition occurs during AIGER encoding.\label{fig:workflow}}
\end{figure*}

We describe \texttt{ltlsynt}'s general approach in Figure~\ref{fig:workflow}.
Let us ignore the \emph{decompose} box and the bypass above the blue area for now.
The main step of the synthesis process is to convert the LTL specification
constraining the input and output signals into a deterministic parity
automaton (DPA) where transitions labeled by Boolean combinations of input
signals are followed by transitions labeled by Boolean combinations of output
signals.  This step is shown in the blue-colored box and discussed in Section~\ref{sec:pipelines}.
This DPA uses a transition-based max-odd parity acceptance. 
We then interpret this DPA as a game between two players (the
\emph{environment} playing the input signals and the \emph{controller}
playing the output signals) and search a winning strategy for the
\emph{controller} using a transition-based version of Zielonka's
algorithm~\cite{zielonka.98.tcs}, then encode this strategy as an
AIGER circuit.

\subsection{Determinization pipelines}\label{sec:pipelines}

The algorithm used by \texttt{ltlsynt} to convert the LTL input into a DPA suitable for game
solving depends on the \texttt{-{}-algo} command-line argument. The first two
options, \texttt{ds} and \texttt{sd}, correspond to pipelines that appeared
in \texttt{ltlsynt}'s very first release. With \texttt{-{}-algo=ds},
LTL inputs are first converted to non-deterministic Büchi automata,
then determinized to DPA using a variant of Safra. At this point,
transitions are labeled by a mix of input and output signals, so
transitions of the form \begin{tikzpicture}[smallautomaton,baseline=-1mm,font=\small]
  \node[state] (x) at (0,0) {};
  \node[state] (y) at (2.5,0) {};
  \path[->] (x) edge[bend right=10] node[overlay,pos=.47,above=-1pt]{$i_1{\land} i_2{\land} o_1 {\land} o_2$} (y);
\end{tikzpicture}
 are split into
\begin{tikzpicture}[smallautomaton,baseline=-1mm,font=\small]
  \node[state] (x) at (0,0) {};
  \node[state] (y) at (1.3,0) {};
  \node[state] (z) at (2.6,0) {};
  \path[->] (x) edge[bend right=10] node[overlay,above=-1pt]{$i_1{\land}i_2$} (y);
  \path[->] (y) edge[bend right=10] node[overlay,above=-1pt]{$o_1{\land}o_2$} (z);
\end{tikzpicture}.  To preserve determinism, we ensure
that multiple transitions sharing the same inputs end up sharing the same
new intermediate state. In the pipeline \texttt{-{}-algo=sd}, we perform
this split before determinizing the automata. Intuitively, this choice
may be explained by realizing that the determinization function, in order
to compute the successors of a given state, has to consider all compatible assignments
of the atomic propositions used by transitions leaving said state: in
\texttt{ds}, there might be up to $2^{|I|+|O|}$
assignments to consider, whereas in \texttt{sd} a given state has
either $2^{|I|}$ or $2^{|O|}$ successors at most.

Option \texttt{-{}-algo=lar.old} in Spot 2.9 used to be called
\texttt{-{}-algo=lar} in versions 2.7 and 2.8, and relies on Spot's
ability to translate LTL formulas into automata with Emerson-Lei
acceptance condition (i.e., any acceptance condition).  To do so, this
algorithm decomposes the input LTL formula on Boolean operators,
translates sub-formulas into deterministic automata (possibly using
algorithms specialized for a particular class of formulas), recombines
the resulting automata using synchronous products, then applies the
relevant Boolean operations on the acceptance conditions.  If we are
lucky enough, we may avoid Safra-based determinization entirely.
However, the resulting deterministic automaton may feature some
arbitrary conditions that have yet to be paritized.  Therefore, we use
a transition-based adaptation of the \emph{state appearance record}
algorithm, typically used to convert state-based Muller acceptance to
state-based parity.  This option was named LAR as a reference to the
\emph{latest appearance record} family of algorithms to which SAR
belongs (some variants of SAR are often called LAR).  We called our
variant CAR, for \emph{color appearance record} as it tracks only the
colors but neither the states nor the transitions.

The CAR implementation in Spot 2.7 featured an optimization that
reduced the number of colors tracked by computing the classes of
symmetric colors in the acceptance condition (two colors are symmetric
if swapping them in the acceptance formula results in an equivalent
formula).  The intent was to keep track of a smaller number of
acceptance classes instead of colors, but this optimization was found
to be incorrect during SYNTCOMP'19. This optimization was removed from
Spot 2.8 for correctness sake, then replaced by many new optimizations
in Spot 2.9~\cite{renkin.20.atva}.

Option \texttt{-{}-algo=lar} now triggers the new implementation of
the paritization procedure~\cite{renkin.20.atva}.  It combines CAR (a
generic transformation to parity) with IAR (a conversion of Rabin-like
or Streett-like acceptance conditions to parity) as well as a
partial-degeneralization (in order to reduce conjunctions of
$\mathsf{Inf}$ or disjunctions of $\mathsf{Fin}$ that occur in the
acceptance condition to a single term, as intended by the original
symmetry-based optimization) and multiple simplifications of the
acceptance conditions.  All these transformations are performed on
each SCC separately, and it may for instance happen that one SCC is
paritized using CAR while another SCC is partially degeneralized to
produce an acceptance condition that can be paritized with IAR.  Our
benchmarks performed on data from SYNTCOMP'17 suggest that the option
\texttt{-{}-algo=lar} often produces significantly smaller DPAs than
\texttt{-{}-algo=lar.old}~\cite{renkin.20.atva}.

A new option available since Spot 2.9.1 is
\texttt{-{}-algo=ps}.  This is a close variant of
\texttt{-{}-algo=ds}, that relies on the translation code that
powers \texttt{ltl2tgba -P -D} to obtain a DPA.
This procedure splits the top-level LTL formulas on Boolean operators
in order to translate subformulas corresponding to obligations
formulas separately.  The remaining subformulas are separately
translated to NBA, determinized using Safra if needed, then
combined back with the obligation part.
Our preliminary experiments showed this option to be inferior to the
other methods, and since we had to pick three configurations for this
year's competition, we excluded this procedure.  In the future it
could be improved by tagging the subformulas based on their
corresponding acceptance conditions, as performed by
Strix~\cite{luttenberger.20.ai}.

\subsection{Various optimizations}

We now discuss other optimizations introduced since 2018.

\smallskip
\noindent\textbf{LTL decomposition}
If the input specification can be seen as a conjunction
$\psi_1\land\psi_2\land\ldots\land\psi_n$ of subformulas with disjoint
output variables, then a strategy for each $\psi_i$ can be computed
separately, as suggested by \citet{finkbeiner.21.nfm}.  Unlike in their
experiments, we recompose the different strategies during the AIGER
encoding, in case they may share gates.

\smallskip
\noindent\textbf{Translation}
Since Spot 2.7, the LTL translation engine (that stands behind the
``translate to $xx$A'' boxes in Figure~\ref{fig:workflow}) learned to
split the input formula on Boolean operators in order to separately translate
parts to automata  then combine these to produce the desired
result.  This is similar to the process used by the \texttt{delag}
tool~\cite{muller.17.gandalf}, but we use slightly improved algorithms.
Extracting \emph{obligations} subformulas is beneficial because those can be
converted to minimal weak deterministic automata~\cite{dax.07.atva}.
Subformulas of the form $\mathsf{GF}(\mathit{guarantee})$ or
$\mathsf{FG}(\mathit{safety})$ can be converted to DBA or DCA using
dedicated algorithms (our implementation is a crossover between two
different works~\cite{esparza.18.lics,muller.17.gandalf}).  Finally,
the products combining the resulting automata handle
weak-automata and suspendable properties~\cite{babiak.13.spin}
specifically. The heuristics used depend on the type of automata
to produce.  For instance, in order to generate NBA or DBA, we
only split the LTL formula on conjunctions.  The post-Spot-2.9 version
submitted to SYNTCOMP also deals with xor and equivalence
operators while converting to DELA (following
Strix's footsteps~\cite{luttenberger.20.ai}).
%

\smallskip
\noindent\textbf{Parity minimization}
Spot 2.8 features a function that minimizes the number of colors
in a DPA~\cite{carton.99.ita}, now called in
\texttt{ltlsynt} once a DPA is produced, before merging states with
identical successors.

\smallskip
\noindent\textbf{Split --- from automata to arenas}
The split operation described above transforms an automaton into a
two-player arena. Even though this step is merely a technicality,
benchmarks on Spot 2.9 have shown that it can consume up to 20\% of
the total run time. In the submitted version, this process has been optimized
thanks to caching operations, as labels are often shared among multiple
transitions. Moreover, the number of edges and states has been reduced
by sharing the introduced intermediate states.

\smallskip
\noindent\textbf{Solving the game}
Since 2020, the parity game solver of \texttt{ltlsynt} is a
transition-based adaptation of the one
from~\citet{vandijk.18.tacas}. It supports (non-recursive) SCC
decomposition, parity compression (a.k.a.  priority compression) and
detection of sub-arenas having a single parity.  In the majority of
the SYNTCOMP benchmarks, solving the parity game is not the bottleneck
of \texttt{ltlsynt}, but nonetheless remains a crucial step as it also
determines the strategy which directly influences the size of the
resulting AIGER circuit.

\smallskip
\noindent\textbf{Bypassing the game}
For inputs of the form
$\LTLG(b_1) \land (\varphi \leftrightarrow \LTLG\F b_2)$, where $b_1$ is a
synthetizable Boolean formula, $\varphi$ is a DBA-realizable property
(a.k.a. recurrence) using only input variables, and $b_2$ is a Boolean
formula using only output variables, a strategy can be constructed
directly from the DBA by ``anding'' each transition with:
$b_1\land\lnot b_2$ if the transition \emph{can} belong to a rejecting
cycle, $b_1\land b_2$ if it \emph{always} belong to an accepting
cycle, or $b_1$ if it cannot belong to any cycle.  (If a ``false'' transition
is created, the formula is unrealizable.)

\smallskip
\noindent\textbf{Strategy simplification}
The winning strategy of a game can be seen an incompletely specified
Mealy machine (ISMM): the value of output variables might be
unspecified when it does not matter.  We implement two algorithms for
the simplification of such ISMM.  One is a variant of Spot's
simulation-based reduction based on BDD
signatures~\citep{babiak.13.spin}, where to states whose signature are
equivalent up-to-unspecified outputs, can be merged.  A second is our own
reimplementation of MeMin's SAT-based minimization algorithm for
ISMM~\citep{abel.15.iccad}.

\smallskip
\noindent\textbf{Optimizing the output circuit}
For the synthesis track submission, the AIGER output of
\texttt{ltlsynt} is run through \texttt{abc} for
simplification~\cite{brayton.10.cav}.  This is done in the driver
script for \texttt{starexec}, not by \texttt{ltlsynt} itself.

\bibliographystyle{abbrvnat}
\bibliography{mc}

\end{document}